\documentclass{aastex631} 

\newcommand{\RomanNumeralCaps}[1]{\MakeUppercase{\romannumeral #1}}
\usepackage{booktabs,multirow}
\usepackage{amsmath}
\usepackage{xcolor}

\shorttitle{Improved AISFMs and Their Release}
\shortauthors{Jeong et al.}

\graphicspath{{./}{figures/}}

\begin{document}

\title{Improved AI-generated Solar Farside Magnetograms by STEREO and SDO Data Sets and Their Release}

\author[0000-0003-4616-947X]{Hyun-Jin Jeong}
\affiliation{School of Space Research, Kyung Hee University, Yongin, 17104, Republic of Korea}

\author[0000-0001-6216-6944]{Yong-Jae Moon}
\affiliation{School of Space Research, Kyung Hee University, Yongin, 17104, Republic of Korea}
\affiliation{Department of Astronomy and Space Science, College of Applied Science, Kyung Hee University, Yongin, 17104, Republic of Korea}

\author[0000-0003-0969-286X]{Eunsu Park}
\affiliation{Space Science Division, Korea Astronomy and Space Science Institute, Daejeon, 34055, Republic of Korea}

\author[0000-0002-9300-8073]{Harim Lee}
\affiliation{Department of Astronomy and Space Science, College of Applied Science, Kyung Hee University, Yongin, 17104, Republic of Korea}

\author[0000-0002-0230-4417]{Ji-Hye Baek}
\affiliation{Technology Center for Astronomy and Space Science, Korea Astronomy and Space Science Institute, Daejeon, 34055, Republic of Korea}
\affiliation{Space Science Division, Korea Astronomy and Space Science Institute, Daejeon, 34055, Republic of Korea}

\correspondingauthor{Yong-Jae Moon}
\email{moonyj@khu.ac.kr}


\begin{abstract}

Here we greatly improve Artificial Intelligence (AI)-generated solar farside magnetograms using data sets of Solar Terrestrial Relations Observatory (STEREO) and Solar Dynamics Observatory (SDO).
We modify our previous deep learning model and configuration of input data sets to generate more realistic magnetograms than before.
First, our model, which is called Pix2PixCC, uses updated objective functions which include correlation coefficients (CCs) between the real and generated data.
Second, we construct input data sets of our model: solar farside STEREO extreme ultraviolet (EUV) observations together with nearest frontside SDO data pairs of EUV observations and magnetograms.
We expect that the frontside data pairs provide the historic information of magnetic field polarity distributions.
We demonstrate that magnetic field distributions generated by our model are more consistent with the real ones than before in view of several metrics.
The averaged pixel-to-pixel CC for full disk, active regions, and quiet regions between real and AI-generated magnetograms with 8 by 8 binning are 0.88, 0.91, and 0.70, respectively.
Total unsigned magnetic flux and net magnetic flux of the AI-generated magnetograms are consistent with those of real ones for test data sets. 
It is interesting to note that our farside magnetograms produce consistent polar field strengths and magnetic field polarities with those of nearby frontside ones for solar cycle 24 and 25.
Now we can monitor the temporal evolution of active regions using solar farside magnetograms by the model together with the frontside ones. 
Our AI-generated Solar Farside Magnetograms (AISFMs) are now publicly available at \href{http://sdo.kasi.re.kr}{Korean Data Center (KDC) for SDO}.

\end{abstract}


\defcitealias{Jeong2020}{Jeong20}
\defcitealias{Kim2019}{KPL19}


\section{Introduction} \label{sec:intro}

Magnetic fields play a fundamental role in producing solar extreme events, i.e., solar flares and coronal mass ejections \citep{wiegelmann2014, judge2021}.
A series of ground-based and space-borne magnetographs have provided solar magnetic field data to study the field's origin and evolution over the last tens of years \citep{pietarila2013}.
The solar magnetograph is an instrument producing a map of magnetic field strength and/or direction on the Sun, and the map is called magnetogram \citep{babcock1953}.
Helioseismic and Magnetic Imager \citep[HMI;][]{Schou2012} on board SDO, which is in geosynchronous orbit, has provided high-resolution magnetograms of the entire solar disk.
Recently, Polarimetric and Helioseismic Imager \citep[PHI;][]{solanki2020} on board Solar Orbiter starts obtaining data of photospheric fields from outside the Sun-Earth line \citep{muller2020}.

Before the Solar Orbiter mission, twin STEREO spacecrafts provided the first stereoscopic view of the Sun drifting ahead and behind the Earth's orbit \citep{Kaiser2008}.
The STEREO Ahead (A) and Behind (B) were launched in 2006 and offered a complete 360 degree view of the entire Sun with the frontside observations.
The STEREO data, together with the SDO ones, have been widely used to study the solar atmospheric phenomena in three dimensions \citep{sterling2012,caplan2016,zhou2021}.
However, because they did not have magnetograph, there was a limit to studying magnetic activities from the frontside to the farside of the Sun during the STEREO era.

Kim, Park, Lee et al. (\citeyear{Kim2019}; hereafter \citetalias{Kim2019}) generated solar farside magnetograms from the STEREO/Extreme UltraViolet Imager \citep[EUVI;][]{Howard2008} 304 {\AA} observations using deep learning.
\citetalias{Kim2019} applied Pix2Pix model \citep{Isola2017} which is a widely used deep learning model in image translation tasks.
They trained and evaluated the model with pairs of solar frontside SDO/Atmospheric Imaging Assembly \citep[AIA;][]{Lemen2012} 304 {\AA} observations and SDO/HMI line-of-sight (LOS) 720s magnetograms.
Results of \citetalias{Kim2019} showed that the farside magnetograms could be used to monitor the temporal evolution of active regions.
However they set the upper and lower saturation limits of the field strength at $\pm 100$ Gauss, because their model worked well with proper byte scaling \citep{park2021}.
Their model well generated the distributions and shapes of the active regions, but it was hard to produce original scale magnetic fluxes \citep{liu2021}.
Jeong et al. (\citeyear{Jeong2020}; hereafter \citetalias{Jeong2020}) improved the AI-generated magnetograms using an upgraded deep learning approach with $\pm 3000$ Gauss dynamic range based on the Pix2PixHD model \citep{Wang2018} and multi-channel SDO/AIA images of 171, 195, and 304 {\AA} for the model input.
\citetalias{Jeong2020} showed that their results could reproduce strong magnetic fluxes and distribution of polarities for not only active regions (ARs), but also quiet regions (QRs).
They applied the AI-generated farside ones to a part of the boundary conditions for the extrapolation of coronal magnetic fields.
Results of the application were much more consistent with coronal farside EUV observations than those of the conventional method.

In the present study we generate more accurate solar farside magnetograms than those of \citetalias{Kim2019} and \citetalias{Jeong2020}.
For this we make an upgraded model including a CC-based objective with additional input data: not only farside STEREO EUV images but also frontside data pairs of SDO/AIA EUV images and HMI magnetograms as reference information.
In this paper, we call the farside data generated by \citetalias{Kim2019} AISFM 1.0, the data generated by \citetalias{Jeong2020} AISFM 2.0, and our data AISFM 3.0, respectively.
We describe detailed structure of our model in Section \ref{sec:Model}, and our data configurations in Section \ref{sec:Data}. 
We show our evaluation results of the model trained with the frontside evaluation data sets in Section \ref{sec:Evaluation}.
Then we generate AISFMs by the model from the corresponding images of STEREO A (or B) and the frontside reference data pairs, and show the results in Section \ref{sec:Generation}.
We release the AISFMs 3.0, and describe the data in Section \ref{sec:Release}. 
We conclude our study in Section \ref{sec:Conclusion}.


\section{Deep Learning Model} \label{sec:Model}

\begin{figure*}[t]
\includegraphics[scale=0.48]{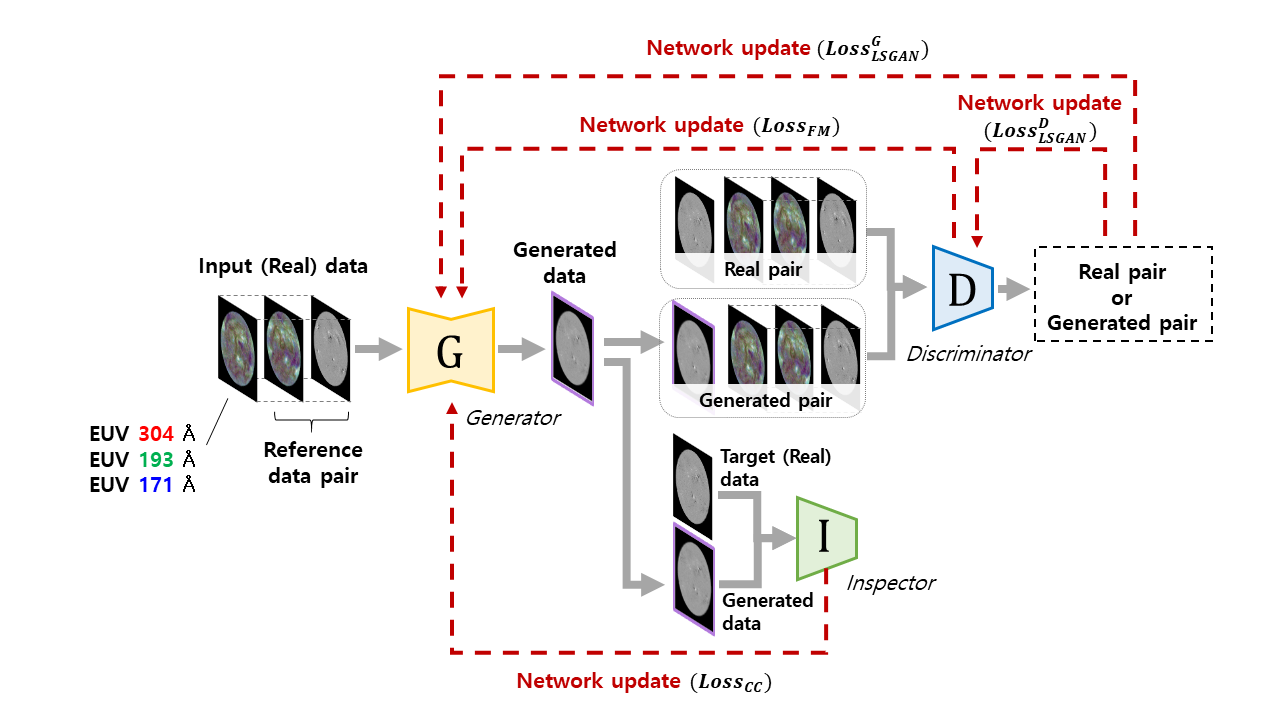}
\centering
\caption{
Flowchart and structures of Pix2PixCC model. $G$, $D$, and $I$ are generator, discriminator, and inspector, respectively.
The generator produces target-like data from input data.
When we train the model, input data are SDO/AIA EUV 304, 193, and 171 {\AA} images and a reference data pair.
The reference data pair is composed of the three SDO EUV images and the corresponding SDO/HMI magnetogram which are the nearest available ones.
The discriminator trains for distinguishing between the real pair and generated pair. 
The real pair consists of an input data and a target data, and the generated pair consists of an input data and a generated data.
The inspector computes CCs between the target data and generated data.
The generator and discriminator are updated from the losses calculated from the inspector and discriminator.
}
\label{f01}
\end{figure*}

We use a deep learning model called Pix2PixCC model to generate solar farside magnetograms from the farside EUV observations and reference data pairs.
Figure \ref{f01} shows the main structure of our model.
The model consists of three major components: a generator ($G$), a discriminator ($D$), and an inspector ($I$).
The generator is a generative network, and tries to produce target-like data from input with the help of the discriminator and the inspector.
The discriminator is a discriminative network, which attempts to distinguish between the more realistic pair between a real pair and a generated pair.
The real pair consists of an input data and a target data.
The generated pair consists of an input data and an output from the generator.
The generator gets updated with objectives from the discriminator, and tries to generate the best outputs to fool the discriminator. 
The inspector computes CCs between the target data and output from the generator to produce realistic values for the generated ones.

%

The generator and discriminator are multi-layer networks.
The multiple layers of the generative network are composed of several convolutional and transposed-convolutional filters of which parameters are updated during the model training process.
Briefly, the convolutional filters try to extract features automatically from the input data, and the transposed-convolutional filters attempt to reconstruct outputs from the extracted features.
For the detailed function of the filters in the network, refer to \cite{goodfellow2016} and \cite{buduma2017}.
The discriminative network consists of several convolution layers.
Each convolution layer generates a feature map based on input.
Our model has a Feature Matching (FM) loss ($Loss_{FM}$), which is an objective function to optimize the parameters of the generator.
The FM loss is to minimize the absolute difference between the feature maps of the real and generated pair from multiple layers of the discriminator.
It is more effective for a large dynamic range data than the loss function derived from the absolute difference between the target and generated data directly \citep{rana2019, marnerides2021}.
The objective function $Loss_{FM}$ is given by
\begin{equation}\label{Lfm}
{Loss}_{FM}(G,D) = \sum_{i=1}^{T} \frac{1}{N_{i}} \Big\|D^{(i)} (x,y)-D^{(i)} (x,G(x)) \Big\|,
\end{equation}
where $x$, $y$, $T$, $i$, and $N_{i}$ are input data, target data, the total number of layers, the serial number of the layers, and the number of elements in output feature maps of each layer, respectively.
$G(x)$ means output data from the generator.
$D^{(i)}$ denotes the $i$th-layer feature extractor of the discriminator.

Our networks use Least Squares Generative Adversarial Network (LSGAN) losses \citep{mao2017}.
The LSGAN losses update the generator ($Loss^{G}_{LSGAN}$) and the discriminator ($Loss^{D}_{LSGAN}$), and are obtained by
\begin{equation}\label{LSGAN}
\begin{aligned}
&{Loss}_{LSGAN}^{G}(G,D) = \frac{1}{2}\Big(D(x,G(x))-1\Big)^2 \\
&{Loss}_{LSGAN}^{D}(G,D) = \frac{1}{2}\Big(D(x,y)-1\Big)^2 + \frac{1}{2}\Big(D(x,G(x))\Big)^2, 
\end{aligned}
\end{equation}
where $D(x,y)$ and $D(x,G(x))$ are probabilities in the range of 0 (generated) to 1 (real) at the end of discriminator from real pair and generated pair, respectively. 
While the generator tries to minimize the $Loss^{G}_{LSGAN}$, and discriminator tries to minimize $Loss^{D}_{LSGAN}$.
The competition between the generator and the discriminator contributes to generate realistic data.
Performance of the adversarial objectives have been well demonstrated in image-to-image translation tasks for solar data\citep{park2019, shin2020, lim2021, son2021}.

In order for stable training of the generator, we use an additional objective function called Correlation Coefficient (CC) loss ($Loss_{CC}$).
It is known that the CC-based loss function has resulted in better performance than error-based loss functions: mean squared error, mean absolute error, etc \citep{vallejos2020, atmaja2021}.
We use Lin's concordance CC, which takes bios into Pearson's CC \citep{lawrence1989}.
The concordance CC is commonly used to assess the reproducibility evaluating the degree to which pairs of data fall on the $45^{\circ}$ line through the origin. 
The range of concordance CC is from -1 (perfect disagreement) to 1 (perfect agreement).
The inspector compute the CC loss with multi-scale target and generated data.
The function of $Loss_{CC}$, which maximizes the agreement between target and generated data, is defined as
\begin{equation}\label{CC}
\begin{aligned}
&{Loss}_{CC}(G) = \sum_{i=0}^{T} \frac{1}{T+1} \Big(1- CC_{i}(y, G(x))\Big), 
\end{aligned}
\end{equation}
where T and i are the total number of downsampling by a factor of 2 and the serial number of the downsampling, respectively.
$CC_{i}$ means the CC value between the $2^{i}$ times downsampled target and AI-generated data.
The average of the CC values from multi-scale target and generated data helps the model to optimize the network parameters.
In addition, with the help of $Loss_{CC}$, we do not impose artificial saturation limits on our model.

Our final objectives are as follows:
\begin{equation}\label{Lfinal}
\begin{aligned}
&\min_{G}\Big(\lambda_{1} {Loss}^{G}_{LSGAN}(G,D) + \lambda_{2} {Loss}_{FM}(G,D) + \lambda_{3} {Loss}_{CC}(G)\Big) \\
&\min_{D}\Big({Loss}^{D}_{LSGAN}(G,D)\Big),
\end{aligned}
\end{equation}
where $\lambda_{1}$, $\lambda_{2}$ and $\lambda_{3}$ are hyperparameters which control the importance of $Loss^{G}_{LSGAN}$, $Loss_{FM}$, and $Loss_{CC}$, respectively.
We use 2, 10, 5 for $\lambda_{1}$, $\lambda_{2}$ and $\lambda_{3}$, respectively.

The purpose of the GAN objectives is to generate an answer which is acceptable. 
It is designed to deal with the probability space of the output.
\cite{Wang2018} improved the GAN objectives by incorporating the FM objective to produce stable outputs.
They set that the importance of the FM loss ($\lambda_{2}$) higher than that of the GAN loss ($\lambda_{1}$).
In \citetalias{Jeong2020}, we showed that realistic magnetograms can be produced by the Pix2PixHD model, which uses both FM and GAN losses.
In the present study, we use not only the FM loss but also the CC loss.
The CC objective guides our model to generate the fields balancing positive and negative polarities.
As a result of multiple tests with different values of the importance, we set 5 for the importance of CC loss ($\lambda_{3}$), of which our model shows the best performance in terms of metrics and visual aspects.
The importance of GAN loss in our final objectives is lower than that of Pix2PixHD.
As the training of our model continues, the model gets more updates by the FM and CC loss than the GAN loss.
To minimize the objective functions, we use an adaptive moment estimation \citep[Adam;][]{Kingma2015} optimizer with learning late 0.0002.
We train the model for 1,000,000 iterations, and save the model and AI-generated data from the evaluation inputs at every 10,000th iteration.
We evaluate all the saved models by the metrics and, use the highest scoring model to generate farside magnetograms.
Our codes of the Pix2PixCC are available at \url{https://github.com/JeongHyunJin/Pix2PixCC}, and more details of our model are described in the readme file.
The codes are archived on Zenodo at \url{https://doi.org/10.5281/zenodo.6668849}.


\section{Data sets} \label{sec:Data}

\subsection{Training data sets} \label{sec:TrainDS}

Here we use SDO/AIA EUV 304, 193, and 171 {\AA} images and SDO/HMI LOS magnetograms to train our deep-learning model.
The three EUV passbands correspond to the chromosphere, corona, and upper transition region of the Sun, respectively.
We use multi-channel inputs to generate target magnetograms.
Channel dimensions of the inputs are composed of three EUV passband images and a reference data pair.
The reference data pair is composed of three SDO/AIA images and a SDO/HMI magnetogram which were observed one solar rotation (27.3 days) before.
We expect that the differences between the EUV images and the reference data pair give the model information on how the intensities or distributions of the magnetic fields have changed.

We use pairs of train data sets with 6 hour cadence (at 01, 07, 13, and 19 UT each day) from 2011 January 1 to 2021 June 30.
We select ten months of data per year, excluding the data sets for the evaluation of our model.
The months are shifted by four months.
Among the 2011 data sets, for example, we use data from March to December for the model training, and the remaining data from January to February for the model evaluation.
We use data from 2012 January to April and July to December to train the model, and the remaining data from May to June to evaluate the model.
These data set configurations are given to consider various solar inclination conditions.
The inclination of the solar rotation axis with respect to the ecliptic plane makes different distributions of southern/northern magnetic fields for each month \citep{Pastor2015}.
We take 6437 pairs for the training data sets.
We remove data with poor quality which are flagged by nonzero value of the \textsf{QUALITY} keyword for both AIA and HMI data sets.
Then we align them to have same rotational axis, pixel size of solar radius ($R_{\odot}$), and location of the disk center.
We downsample them from $4096 \times 4096$ to $1024 \times 1024$ for computational capability.
The radius of the Sun is fixed at 450 pixels.
A mask is applied to the area outside 0.998 $R_{\odot}$ from the disk center for minimizing the uncertainty of limb data.
All EUV data number are scaled by median values on the solar disk to calibrate the gradual in-orbit degradation of the AIA instrument \citep{Ugarte2015,Liewer2017}.

\subsection{Evaluation data sets} \label{sec:EvaluDS}

We use the remaining SDO data pairs, which are two months of data per year, to evaluate our model.
Among them, we use the pairs of data sets from 2011 to 2017 to compare with the results of \citetalias{Kim2019} and \citetalias{Jeong2020}.
We take 1342 pairs for the evaluation data sets.
The pre-processing steps of the evaluation data are the same as those of training data.

When we compute objective measures between the target magnetograms and AI-generated data to evaluate our model, we compare the results not only data for full disk, but also data for ARs and QRs.
We select areas of the ARs and the QRs with a size of $128 \times 128$ pixels from the pre-processed target magnetograms with $1024 \times 1024$ pixel resolution.
We don't consider the areas outside 60 degrees from the disk center to exclude limb data with uncertainty.
We compute the total unsigned magnetic flux (TUMF) for the area, moving at intervals of 64 pixels up, down, left, and right from the center of solar disk.
When the TUMF of the area is greater than $5 \times 10^{21}$ Mx, the area is classified as an AR \citep{Waldmeier1955,van2015}. Otherwise, the area are candidates of QR.
The boundaries of all detected areas do not overlap with one another.
In order to balance the number of AR and QR areas, we select up to 3 AR areas for each magnetogram in the order of the largest TUMF, and up to 2 QR areas for each magnetogram in the order of the lowest TUMF.

%
\begin{figure}[t]
\begin{interactive}{animation}{Figure02_Comparison_between_HMI_and_AI_gen.mp4}
\includegraphics[scale=0.6]{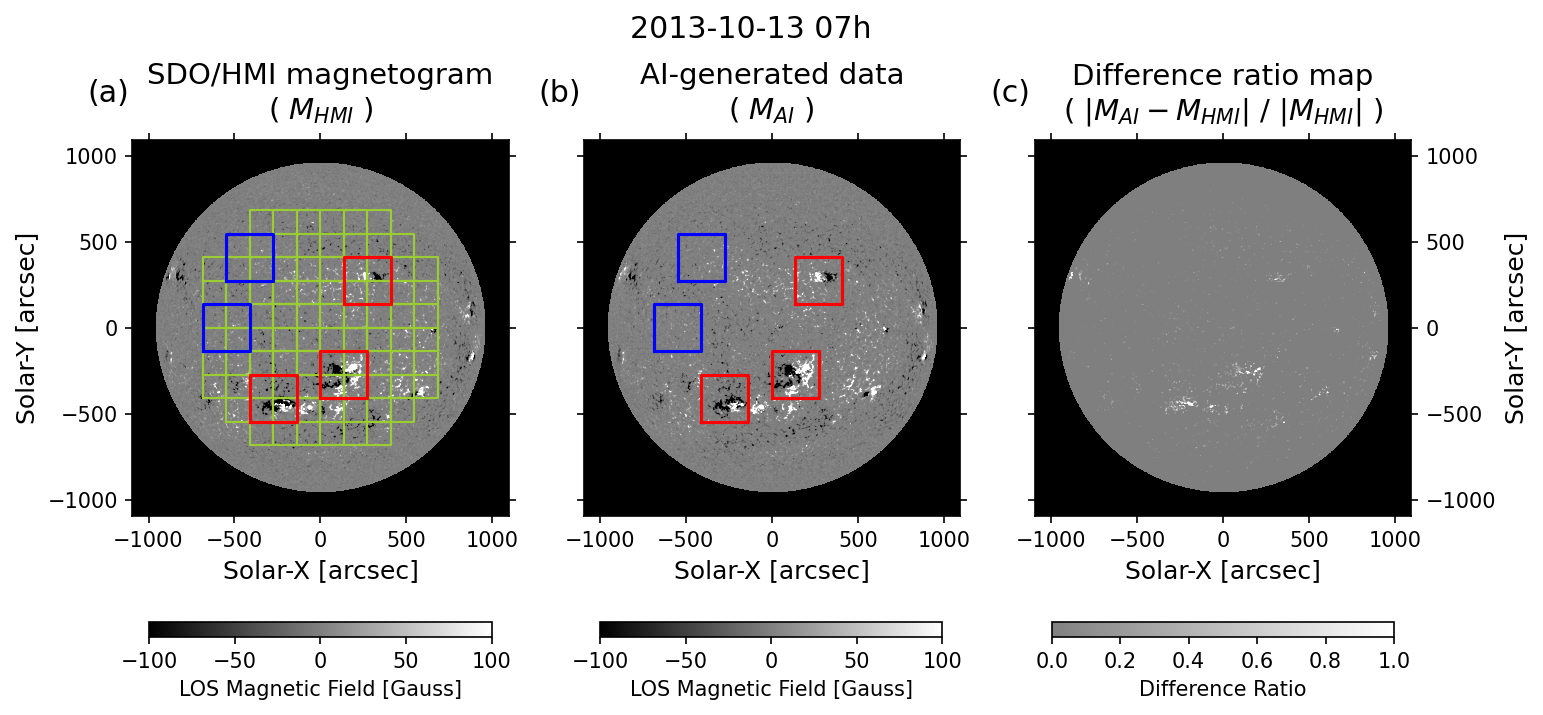}
\centering
\end{interactive}
\caption{
Automatic selected areas of AR (red boxes) and QR (blue boxes) in SDO/HMI magnetogram acquired at 2013 October 13 07:00 UT. 
Polarities of the magnetic fields are displayed as white for positive and black for negative.
The detected areas, each having $128 \times 128$ pixels ($\sim 273 \times 273$ arcsec), are overlaid on AI-generated data corresponding to the real one.
The green grid lines represent 64 pixel ($\sim 136.5$ arcsec) intervals within 60 degrees from the center of solar disk.
(An animation of this figure is available and shows the results from 2013 October 1 to November 30. In the animation, the grid lines are erased for the visual comparison of the results.)
}
\label{f02}
\end{figure}


Figure \ref{f02} shows an example of the selection result on 2013 October 13.
There are three solar active regions near the center of the solar disk.
The green grid lines in Figure \ref{f02}(a) mean the boundaries of the areas where the TUMF is calculated. 
Our method successfully detects the approximate positions of three AR areas (red boxes in Figure \ref{f02}(a)), and two QR areas (blue boxes in Figure \ref{f02}(a)).
Each box is placed in the same position on the AI-generated data as shown in Figure \ref{f02}(b).
Figure \ref{f02}(c) shows a difference ratio map between the SDO/HMI magnetogram and AI-generated data.
To compare the difference of their significant magnetic features, we smooth them out using a method similar to \cite{higgins2011} who used $\pm 70$ Gauss as a minimum threshold and a 2D Gaussian smoothing for the magnetograms; here we take one sigma and window size of $10 \times 10$ pixels.


\subsection{STEREO data sets} \label{sec:StereoDS}

We use solar farside STEREO/EUVI EUV observations and pairs of SDO/AIA EUV images and SDO/HMI magnetograms to generate the farside magnetograms.
The EUV passbands of STEREOs are 304, 195, and 171 {\AA}, which have similar temperatures response to the passbands of the SDO.
We use STEREO data sets with 6 hour cadence (at 00, 06, 12 and 18 UT each day) from 2011 January 1 to 2021 June 30.
Since communications with STEREO B were lost on 2014 October 1, the data from STEREO B are available until that day.
We align, downsample, mask, scale the STEREO EUV images like the SDO EUV ones.
We manually exclude a set of STEREO data with incorrect header information and noise or missing values because of solar flares.
The SDO pairs are data obtained at the frontside ones which are selected by considering the separation angle between STEREO A (or B) and SDO. 
The observation dates of the reference SDO pairs are obtained by
\begin{equation}\label{RefData}
\begin{aligned}
{T}_{SDO} = {T}_{STEREO} - {\Phi}_{STEREO} \times \frac{27.3 \; day}{360^{\circ}}, 
\end{aligned}
\end{equation}
where ${T}_{SDO}$, ${T}_{STEREO}$, and $\Phi_{STEREO}$ are the date of reference SDO pairs, the date of STEREO data sets, and heliographic longitude of the STEREO, respectively. Given these configurations, we expect that the frontside magnetograms give our model the information about the overall magnetic field distribution. And the EUV image sets of the frontside SDO and the farside STEREO are used to give the information about the changes of features on the Sun.

%
\begin{deluxetable*}{lcccccc}
\tablecaption{The average pixel-to-pixel CCs between SDO/HMI magnetograms and AI-generated ones for full disk, ARs, and QRs.
\label{tab:t01}}
\tablehead{
\colhead{} & \multicolumn6c{pixel-to-pixel CC} \\
\colhead{} & \multicolumn3c{$8 \times 8$ binning} \\
\cmidrule(lr){2-4} \cmidrule(lr){5-7}
\colhead{}  &  \colhead{Full Disk} & \colhead{AR} & \colhead{QR}\\
\colhead{}  &  \colhead{(1342)} & \colhead{(2926)} & \colhead{(2684)}
}
\startdata
AISFM 3.0 (Ours) & 0.88 & 0.91 & 0.70 \\
AISFM 2.0 (\citetalias{Jeong2020}) & 0.81 & 0.79 & 0.62 \\
AISFM 1.0 (\citetalias{Kim2019}) & 0.77 & 0.66 & 0.21 \\
\enddata
\tablecomments{
The results of \citetalias{Jeong2020} and \citetalias{Kim2019} are shown for comparison.
}
\end{deluxetable*}

\section{Results and Discussions} \label{sec:Result}


\subsection{Evaluation of Our Deep Learning Model} \label{sec:Evaluation}

We evaluate our deep learning model using the frontside evaluation data sets that we did not use when training the model.
To compare our results with \citetalias{Kim2019} and \citetalias{Jeong2020}, we use Pearson's CC as a measure for the evaluation.
Table \ref{tab:t01} shows the average pixel-to-pixel CCs between SDO/HMI magnetograms and AI-generated ones with a full dynamic range. 
Our model shows that the average pixel-to-pixel CCs after $8 \times 8$ binning are 0.88, 0.91, and 0.70 for 1342 full disk, 2926 ARs, and 2684 QRs, respectively.
These imply that our model improves the generation of magnetograms when compared with results of \citetalias{Kim2019} and \citetalias{Jeong2020}.
In addition, the pixel-to-pixel CCs between the target and our AI-generated data after $4 \times 4$ binning show better results than the CCs between the target and \citetalias{Kim2019}'s ones after $8 \times 8$ binning.
The latitudinal or longitudinal heliographic resolution at the center of the solar disk is approximately 1 degree per pixel after $8 \times 8$ binning.


%
\begin{figure*}[t]
\includegraphics[scale=0.54]{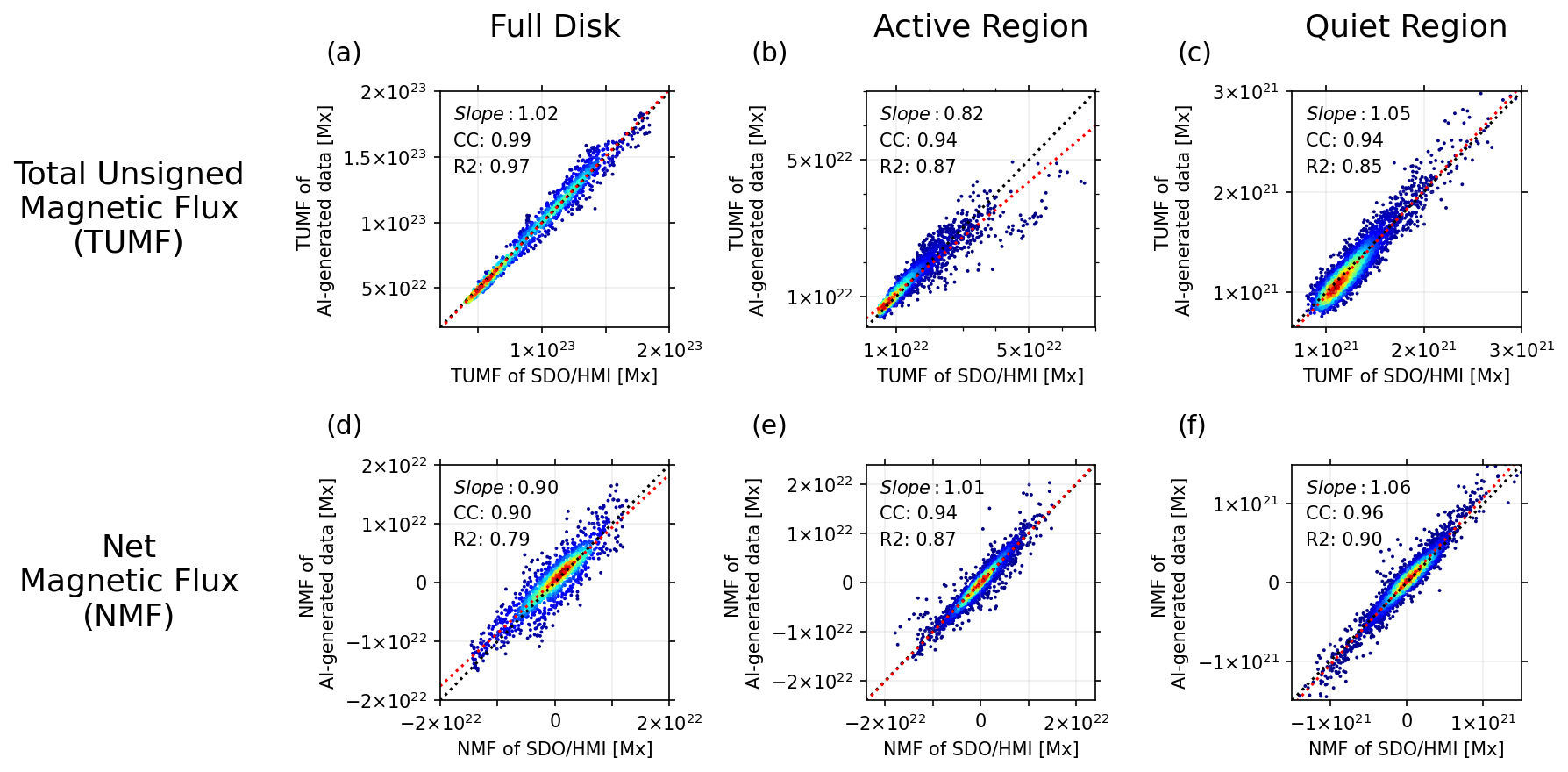}
\centering
\caption{
Two objective measures of comparisons between SDO/HMI magnetograms and AI-generated ones for 1342 full disk, 2926 ARs, and 2684 QRs.
(a), (b), (c) are scatter plots between TUMFs of the SDO/HMI magnetograms and AI-generated ones for the full disk, the ARs, and the QRs, respectively.
(d), (e), (f) are scatter plots between NMFs of the SDO/HMI magnetograms and AI-generated ones for the same data sets, respectively.
}
\label{f03}
\end{figure*}

Figure \ref{f03} shows scatter plots between two objective measures of SDO/HMI magnetograms and AI-generated ones for the same data sets when calculating the average pixel-to-pixel CCs.
We compare TUMF between the target and AI-generated data (Figure \ref{f03}(a) to (c)).
The TUMF CCs are 0.99, 0.94, and 0.94, R2 scores are 0.97, 0.87, and 0.85, and slopes are 1.02, 0.82, and 1.05 for full disk, ARs, and QRs, respectively.
We compare net magnetic flux (NMF) between the real and AI-generated data (Figure \ref{f03}(d) to (f)).
The NMF CCs are 0.90, 0.94, and 0.96, R2 scores are 0.79, 0.87, and 0.90, and slopes are 0.90, 0.94, and 0.96 for full disk, ARs, and QRs, respectively.
Most values of TUMF and NMF fall well on the diagonal line (black dotted line in Figure \ref{f03}) through the origin.
These values support that our model can generate consistent magnetic fluxes.

Figure \ref{f03}(b) shows that our model slightly underestimates TUMF of the strong ARs more than the real ones.
We examine why small portion of ARs show underestimated TUMF.
We find that for some ARs the AI-generated data do not produce strong magnetic fields. In this case, the ARs do not have high intensities at 304, 193, and 171 {\AA} images but have high intensities at 94 {\AA} (Fe \RomanNumeralCaps{18}) and 131 {\AA} (Fe \RomanNumeralCaps{8}, \RomanNumeralCaps{21}) of SDO/AIA. These shorter wavelength channels are characterized by high-temperature emissions \citep{o2010,warren2011}.
We think that the 94 and 131 {\AA} observations are helpful to generate strong magnetic fluxes of the ARs.
However, in this study, we do not use the 94 and 131 {\AA} images to train our model because the STEREOs/EUVI have only filter bands of 171, 195, 284, and 304 {\AA}.

For the ARs, we evaluate our model based on the similarity of the AI-generated magnetograms with the real ones. We use a Structural SIMilarity (SSIM) method, which is widely used to measure the degree of similarity and consider measurements of luminance, contrast, and structure between two images. The SSIM produces a value between 0 and 1. The maximum value of 1 indicates that they show perfectly similar structure, and vice versa. We use dynamic range of the pixel values to compute the SSIM from -1500 Gauss to +1500 Gauss considering our test data sets. Our model shows that average SSIM value for the ARs is 0.74 with the standard deviation of 0.09. After 8 by 8 binning, the average SSIM value is 0.93 with the standard deviation of 0.09.


\subsection{Generation of Solar Farside Magnetograms} \label{sec:Generation}

We generate the farside magnetograms from the STEREO EUV images and the frontside reference data pairs by the model.
The dates of AISFMs for STEREO A (AISFMs A) from 2011 January 1 to 2021 June 30, and the AISFMs for STEREO B (AISFMs B) from 2011 January 10 to 2014 September 27.
In 2011 January, the position of the STEREO A was about $85^{\circ}$ longitude, and  that of the STEREO B was about $-90^{\circ}$ longitude in Stonyhurst heliographic coordinates.
They drift away at a rate of about $22^{\circ}$ per year from the Earth. 

%
\begin{figure*}[t]
\includegraphics[scale=0.6]{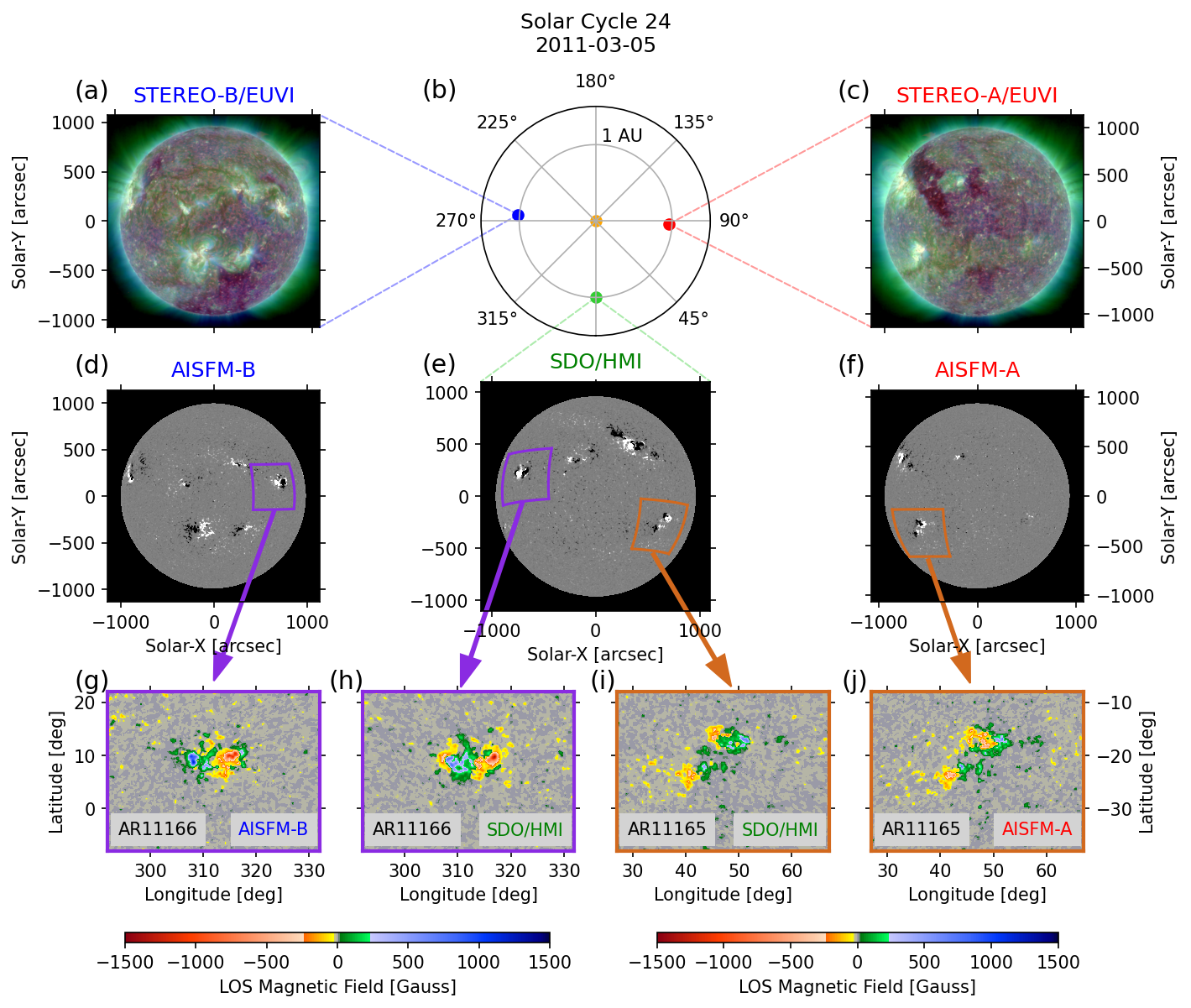}
\centering
\caption{
(a)-(f) Solar frontside SDO/HMI magnetogram, the farside STEREO composite EUV images (red: 304 {\AA}, green: 193 {\AA}, and blue: 171 {\AA}), and AISFMs on 2011 March 5.
The purple and brown boxes represent NOAA AR 11166 and 11165, respectively.
(g)-(j) The two ARs are zoomed and converted from full disk data to heliographic coordinated maps.
The color map of zoomed ARs shows the large dynamic rage values in Gauss.
}
\label{f04}
\end{figure*}

Figure \ref{f04} shows multi-viewpoint data from SDO, and STEREO A and B on 2011 March 5.
The position of STEREO A is about $88^{\circ}$ heliographic longitude near the west limb and the position of STEREO B is about $265^{\circ}$ heliographic longitude near the east limb of the solar frontside (Figure \ref{f04}(b)).
We select a NOAA AR 11165 to the west, and a NOAA AR 11166 to the east of the solar disk from the frontside SDO/HMI magnetogram (Figure \ref{f04}(e)).
The AR 11165 is observed by the STEREO A (Figure \ref{f04}(c)) and the AISFM A (Figure \ref{f04}(f)).
TUMF of the AR 11165 from the SDO/HMI magnetogram in Figure \ref{f04}(i) is about $1.85 \times 10^{22}$ Mx and one from AISFM A in Figure \ref{f04}(j) is about $1.74 \times 10^{22}$ Mx.
The AR 11166 is observed by the STEREO B (Figure \ref{f04}(a)) and the AISFM B (Figure \ref{f04}(d)).
TUMF of the AR11166 from the SDO/HMI magnetogram in Figure \ref{f04}(h) is $2.50 \times 10^{22}$ Mx, and one from AISFM B in Figure \ref{f04}(g) is $2.71 \times 10^{22}$ Mx.
The TUMF of the ARs from our AISFMs are consistent with those of the real one, and distributions of magnetic fields look like the real one.

%
\begin{figure*}[t]
\includegraphics[scale=0.58]{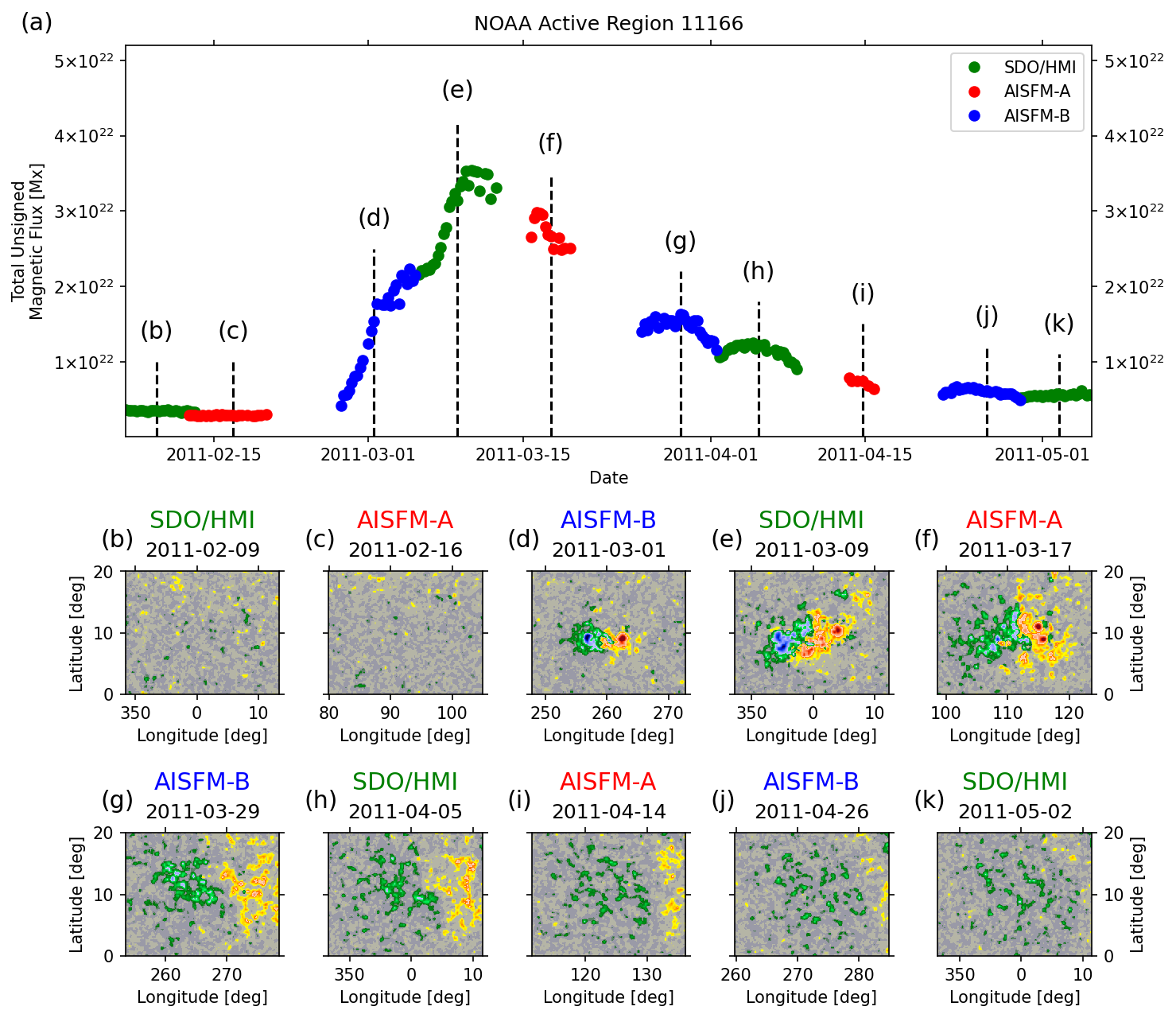}
\centering
\caption{
(a) A temporal variation of TUMF in the NOAA AR 11166 from 2011 February 1 to May 7.
Green, red, blue dots are TUMFs from  SDO/HMI, AISFM A, and AISFM B, respectively.
(b)-(k) A series of magnetograms tracking the AR over three solar rotations.
The magnetograms are converted from full disk data to heliographic coordinated maps.
The color map of magnetic fields is as shown in Figure \ref{f04}(g)-(j).  
}
\label{f05}
\end{figure*}

Figure \ref{f05} shows the temporal evolution of AR 11166, which is shown in Figure \ref{f04}(g)-(h).
We track the AR for three solar rotations at a Carrington rotation rate.
We calculate the TUMF for each area including the AR when the HMI or AISFMs are available, and the results are shown in Figure \ref{f05}(a).
We consider the ARs within 60 degrees from the disk center of the HMI or AISFMs.
One impressive thing is that the TUMFs from the SDO/HMI and those from our AISFMs between solar frontside and farside are smoothly overlapped, demonstrating that it is possible to monitor the change in magnetic flux quantitatively using our method.  
Figure \ref{f05}(b)-(k) shows magnetograms of the tracked AR.
Combining SDO/HMI magnetograms and AISFMs makes it possible to continuously monitor the evolution of magnetic field distribution over the solar surface.

%
\begin{figure}
\begin{interactive}{animation}{Figure06_AISFM_TemporalEvolution.mp4}
\includegraphics[scale=0.55]{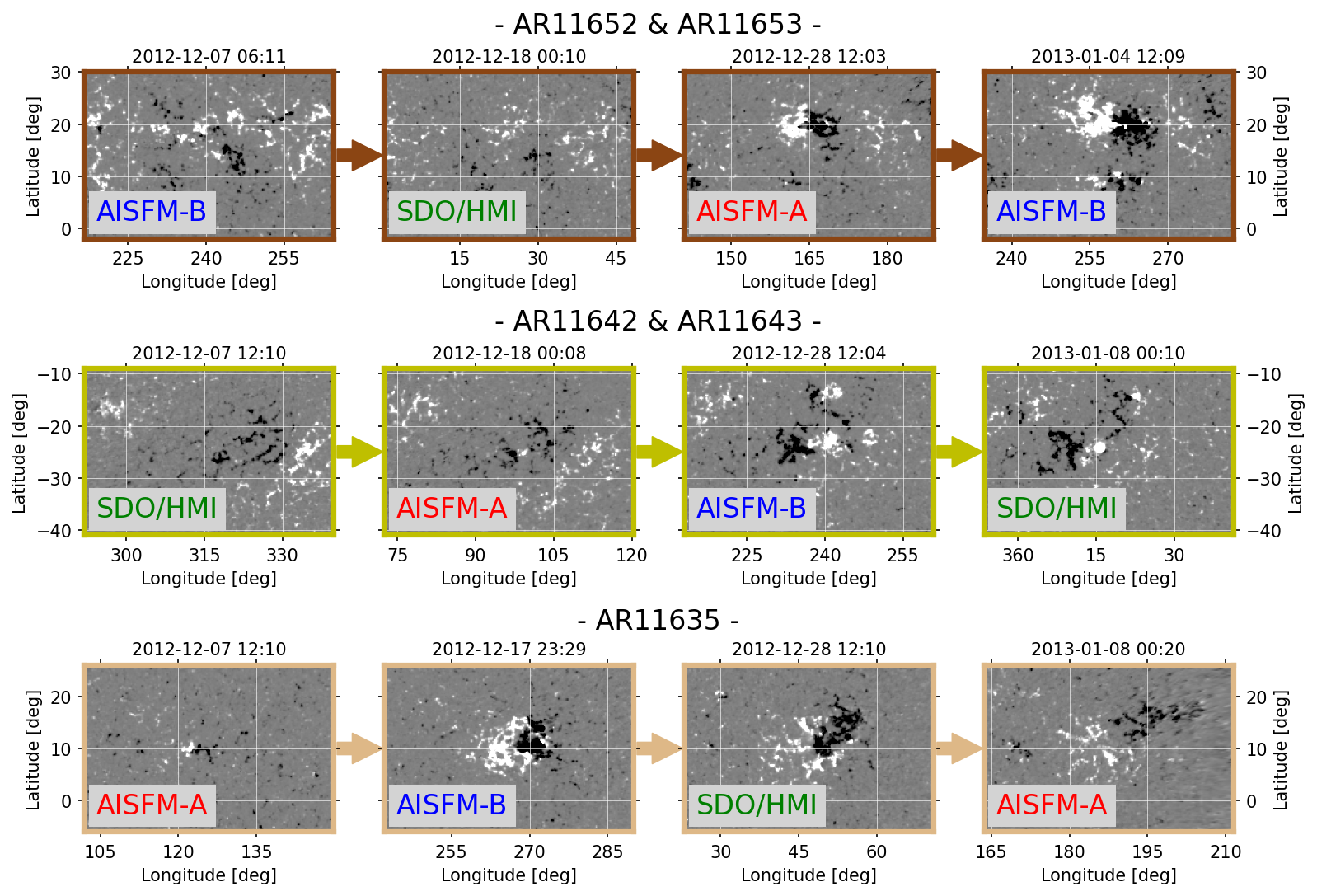}
\centering
\end{interactive}
\caption{
Temporal evolution of the ARs obtained by the SDO/HMI, and AISFM A and B from 2012 December 7 to 2013 January 20. 
Three colored boxes denote the ARs on the solar surface.
(An animation of this figure is available.)
}
\end{figure}

Figure 6 shows tracking of ARs over the solar surface from 2012 December 7 to 2013 January 20. 
The ARs from SDO/HMI, and AISFM A and B are converted from full disk data to heliographic coordinated maps. 
When the data are not available, we replace them with the nearest available ones. 
The position of the STEREO A is about $130^{\circ}$ heliographic longitude, and that of STEREO B is about $230^{\circ}$ heliographic longitude. The AISFM A and B show their consistent growth and decay.

%
\begin{figure*}[t]
\includegraphics[scale=0.6]{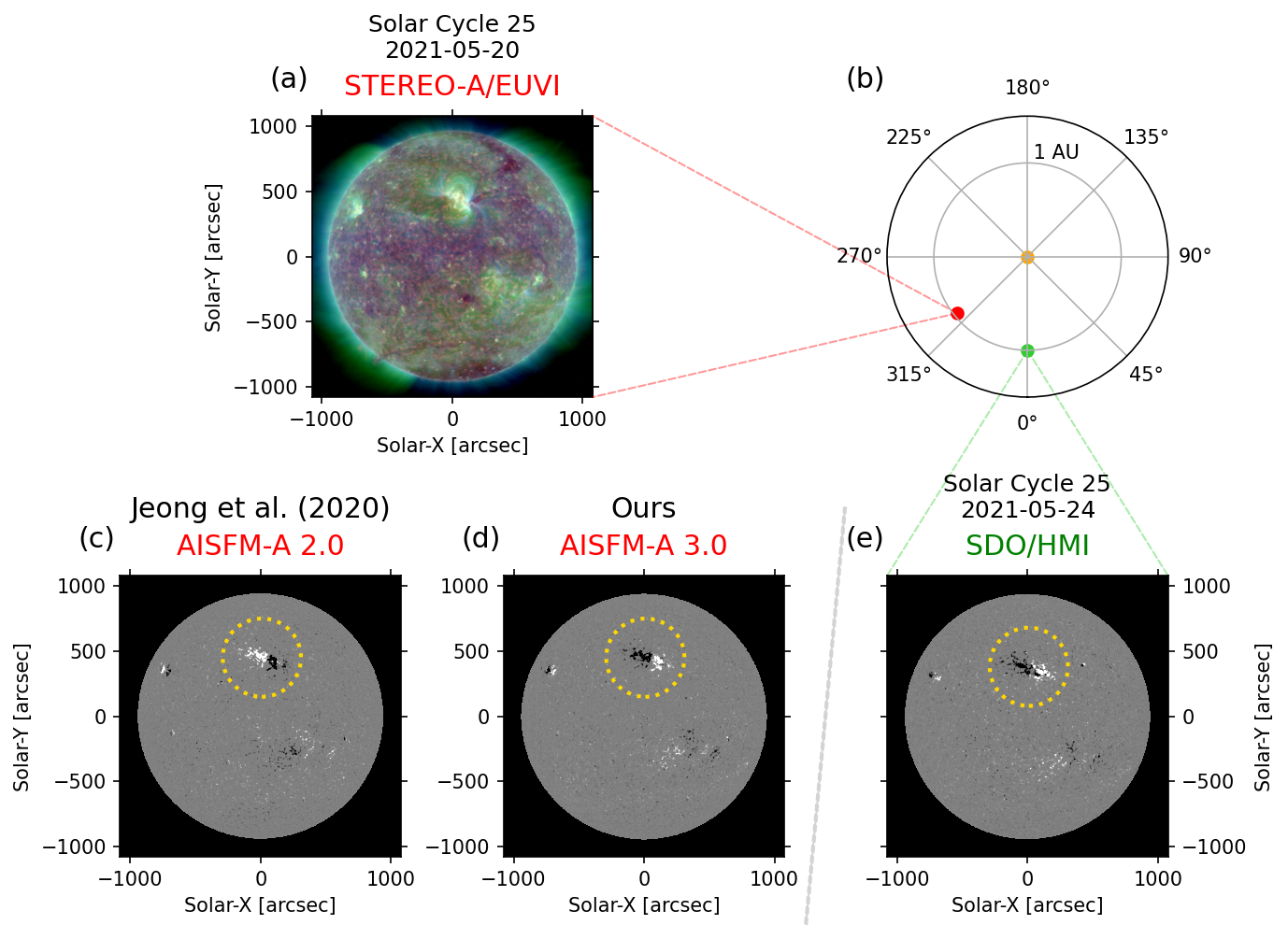}
\centering
\caption{
A Solar farside STEREO A EUV image and AISFMs on 2021 May 20, and the frontside SDO/HMI magnetogram on 2021 May 24.
Yellow dotted circles represent NOAA AR 12824.
A result of \citetalias{Jeong2020} (AISFM A 2.0) is shown together for comparison with our result (AISFM A 3.0).
}
\label{f07}
\end{figure*}

Figure \ref{f07} shows comparisons between a SDO/HMI magnetogram and two AISFMs of solar cycle 25.
When the solar cycle changes, all of solar magnetic field patterns are reversed \citep{hale1925}.
On 2021 May 20, the position of STEREO A is about $309^{\circ}$ heliographic longitude (Figure \ref{f07}(b)).
The magnetic field polarities of our AISFM A (Figure \ref{f07}(d)) are consistent with the ones of a SDO/HMI magnetogram (Figure \ref{f07}(e)), which is obtained at the frontside after 4 days.  
As shown in Figure \ref{f07}(c), AISFM 2.0 cannot produce the reasonable polarities which can be identified from HMI magnetograms.
It is noted that the AISFM 2.0 is generated from the STEREO A EUV observations (Figure \ref{f07}(a)) without reference information from the solar frontside. 
The polarity distributions of AISFM 2.0 are similar to that of solar cycle 24.
We mark NOAA AR 12824 with yellow dotted circles in Figure \ref{f07}(c)-(e).
The AR from the HMI magnetogram shows leading positive and following negative polarities.
Our AISFM 3.0 well represents the polarity distributions of the AR.

%
\begin{figure*}[t]
\includegraphics[scale=0.62]{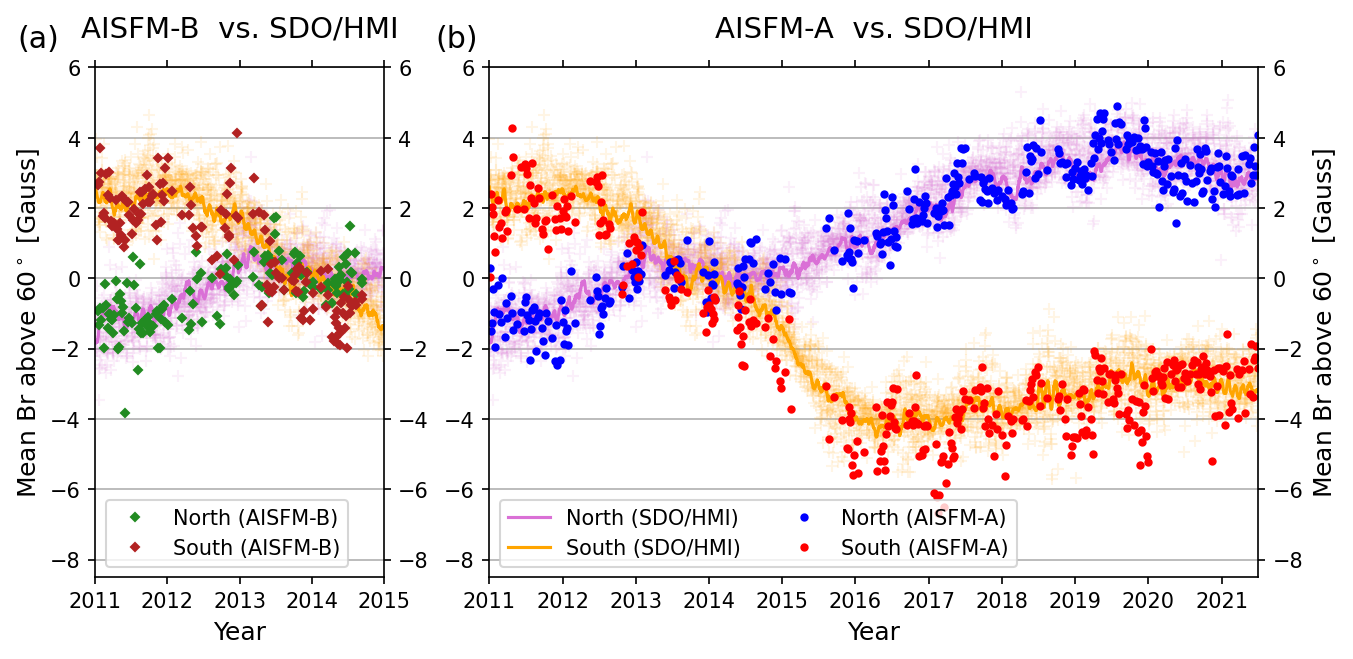}
\centering
\caption{
Comparisons of solar polar field strengths from SDO/HMI and those from AISFM A and B.
Yellow (purple) dots represent the mean radial field strength above $60^{\circ}$ in the south (north) of SDO/HMI magnetograms; the lines represent the smoothed average.
Brown (green) and red (blue) dots represent the mean radial field strength above $60^{\circ}$ in the south (north) of AISFMs A and B, respectively.
}
\label{f08}
\end{figure*}

Figure \ref{f08} shows comparisons between mean polar field strengths from SDO/HMI magnetograms and those from AISFMs A and B.
The results from AISFM A and B are presented at 5 day intervals.
The polar fields computed from our AISFMs 3.0 follow trend of the polar field reversal process shown by the computed results from the SDO/HMI.
We use mean radial fields of the HMI polar field data series, \href{http://jsoc.stanford.edu/ajax/lookdata.html?ds=hmi.meanpf_720s}{$\mathrm{hmi.meanpf\_720s}$}, which are provided from the JSOC.
The LOS magnetic fields are converted to radial field, under the assumption that the actual field vector is radial.
The mean polar field strength is calculated from the values within $\pm 45^{\circ}$ longitude and above $60^{\circ}$ latitude (for details refer to \cite{sun2015}).
We calculate mean polar fields of the AISFMs according to their study, and our results are consistent with their results.

One may ask a question, ``How one technique could find the magnetic polarity distribution from EUV images?''
\citetalias{Kim2019} showed that a deep learning can generate solar farside magnetograms with Hale-patterned active regions being well replicated from the EUV 304 {\AA} images. The pixel values, i.e. intensities, of the EUV images can give our model the distribution of magnetic fields.
\cite{Ugarte2015} showed that integrated 304 {\AA} light curves can be used as a proxy for total unsigned magnetic flux of the AR. Based on these results, several studies tried to generate solar magnetograms from the EUV 304 {\AA} images using deep learning \citep{alshehhi2020, dani2022}. 
\citetalias{Jeong2020} generated more realistic magnetograms using the EUV 304, 193, and 171 {\AA} images. 
The EUV 193 and 171 {\AA} passbands, which corresponds to the corona and upper transition region, are widely used for detection of coronal holes \citep{garton2018, linker2021}.
Distribution of coronal hole are related to that of open flux regions, i.e. unipolar regions \citep{lowder2014}. 
The multi-channel EUV images can give the model information about the distribution of not only the ARs, but also the unipolar regions related to coronal holes. 
Here we use reference solar frontside magnetograms and EUV images to generate the farside magnetograms. 
\cite{hale1919} noted that most leading spots have opposite polarities in opposite hemispheres. 
The Hale's law correctly predicts polarities of the ARs about $90 \%$ of the time \citep{li2018}.
The reference data sets give our model overall magnetic field polarity distributions including the polarities of leading spots.
Based on these arguments, our model successfully generates the farside magnetograms of solar cycle 24 and 25.
However, it may not be exact when the magnetic fields of the ARs do not follow the Hale’s law. 
It is especially difficult to predict the polarity distributions of rapidly emerging ARs, which did not observe at the reference data sets. 
Our model generates magnetograms based on a large amount of iterative training to produce accurate magnetic field distributions from the input data sets. 
If more training data sets with different distribution of magnetic field polarities, various shapes of ARs, appearance and disappearance of ARs are provided, we expect our model to be able to predict more realistic magnetograms.

\section{DATA RELEASE} \label{sec:Release}

Here we first release the AISFMs at the \href{http://sdo.kasi.re.kr}{KDC for SDO} (\url{http://sdo.kasi.re.kr}).
The names of AISFMs 3.0 A and B recorded in the AI-generated data browse are \textsf{aisfm\_v3\_stereo\_a} and \textsf{aisfm\_v3\_stereo\_b}, respectively.
There are 7913 AISFMs A from 2011 January 1 to 2021 June 30, and 2890 AISFMs B from 2011 January 10 to 2014 September 27 with 6 hour cadence.
When the model inputs from the STEREO and the SDO have poor quality data, we do not produce the AISFMs.
The number of AISFMs A for 2015 is smaller than the number of those for other years, because the contact with STEREO A was interrupted as it passed behind the Sun.

The AISFMs 3.0 are saved in the Flexible Image Transport System (FITS) format \citep{pence2010}.
The data have $1024 \times 1024$ pixels and the solar radius is fixed at 450 pixels.
The data outside the solar radius are filled with Not-a-Number (NaN) values like those of SDO/HMI magnetograms.
The data inside the solar radius represent magnetic fields along the LOS to the STEREO A or B.
The coordinate information of the STEREO is stored into the FITS header keywords (see more details in Appendix \ref{sec:header}).
We also provide example codes to understand the AISFMs at \url{https://github.com/JeongHyunJin/AISFM3.0}, and the codes are archived on Zenodo at \url{https://doi.org/10.5281/zenodo.6668571}.

\section{Summary and Conclusion} \label{sec:Conclusion}

In this study we have generated the improved solar farside magnetograms by the STEREO and SDO data sets using a deep learning model.
For this work, we have improved our model including the CC-based objectives and used model inputs the farside STEREO EUV observations together with the frontside SDO data pairs.
We selected 6437 pairs of input and target data sets from 2011 January 1 to 2021 June 30 for the model training.
Targets for the training are SDO/HMI magnetograms.
Inputs for the training consist of the EUV images, and the pairs of EUV images and magnetograms obtained 27.3 days before. 
We have evaluated the model using test data sets not used for training.

The main results of this study are as follows.
First, we improve the AI-generated magnetograms than before.
The average pixel-to-pixel CCs between the SDO/HMI magnetograms and our AI-generated ones after $8 \times 8$ binning are 0.88, 0.91, and 0.70 for full disk, ARs, and QRs, respectively, which are noticeably better than the previous results.
We have good agreements between TUMFs calculated from the SDO/HMI magnetograms and those calculated from the AI-generated ones, and NMFs calculated from the HMI data and those calculated from the AI-generated ones.
Second, we generate more realistic solar farside magnetograms using the STEREO EUV images and the frontside data pairs by the model.
We compare magnetic fields of ARs from the AISFMs and HMI when the STEREOs are about 90 degrees apart from Earth.
Together with the AISFMs and HMI, we can continuously monitor the temporal evolution of TUMF of an AR over three solar rotations.
Third, our model can generate AISFMs of solar cycle 24 and 25, in which data have consistent magnetic field polarities with those of nearby frontside ones.
We show that the temporal variation of mean polar fields calculated from the AISFMs well represents the Sun's magnetic field reversal process. 

Our method has several advantages over the conventional methods.
First, our AISFMs can improve studies using the solar magnetic flux distributions. 
We can track the ARs and study their flux evolution at the solar surface using the AISFMs together with the frontside magnetograms as shown in Figure \ref{f05} (also see \citetalias{Kim2019}). 
Second, we can improve global coronal magnetic field extrapolation from the synchronic maps with our AISFMs. 
In \citetalias{Jeong2020}, we showed that global extrapolations from the synchronic maps with AISFMs were more consistent with EUV observations than those from conventional data in view of the ARs and coronal holes. 
Third, we expect that our AISFMs provide better input data for heliospheric solar wind propagation models such as WSA-ENLIL \citep{arge2000} and EUHFORIA \citep{pomoell2018}. 
We also acknowledge that our method has a couple of limitations.
First, physical quantities based on pixel to pixel distribution of magnetic fields (e.g., neutral line) may not be exact. 
Second, small-scale magnetic field configurations, such as magnetic cancellation features, may not be well produced.

\begin{acknowledgments}

This study uses a large amount of the STEREO and SDO data. 
We appreciate numerous team members who have contributed to the success of the STEREO and SDO mission.
We acknowledge the community efforts devoted to the development of the following open-source packages that were used for this work.
This work was supported by the Korea Astronomy and Space Science Institute (KASI) under the R\&D program (project No. 2022-1-850-05, 2022-1-850-08) supervised by the Ministry of Science and ICT, and the Basic Science Research Program through the NRF funded by the Ministry of Education (NRF-2020R1C1C1003892, NRF-2021R1I1A1A01049615).

\end{acknowledgments}

\software{PyTorch \citep{paszke2019},  
          NumPy \citep{harris2020},  
          Matplotlib \citep{Hunter2007},
          SciPy \citep{scipy2020},
          Astropy \citep{robitaille2013,price2018},  
          SunPy \citep{sunpy_community2020}          
          }

\appendix
\section{FITS header Keywords}\label{sec:header}

AISFMs are stored in FITS files, each with a keyword header containing the information on the data.
The keywords follow World Coordinate System (WCS) conventions for describing the physical coordinate values of the data pixels \citep{greisen2002}, and several ephemeris keywords are provided in Table \ref{tab:ta1}.
Since our data are generated by a deep learning model, not observational one, we store the keyword name not \textsf{OBSERVTRY} (observatory) but \textsf{INPUTDAT} (input data).
The AISFMs are generated based on the features from three EUV images of the STEREO A (or B).
So we record the mean date and time of the three EUV observations in the \textsf{DATE-OBS} keyword.
More detailed information on the model inputs is stored in the \textsf{HISTORY} keyword.

%
\begin{deluxetable}{ccc}
\restartappendixnumbering
\tablewidth{1pt}
\tablecaption{Ephemeris keywords for the AISFMs.
\label{tab:ta1}}
\tablehead{\colhead{Keyword} & \colhead{Description} & \colhead{Format or Unit}}
\startdata
\textsf{INPUTDAT} & Observer of the model input data & STEREO\_A (or \_B) \\
\textsf{DATE-OBS} & Mean date and time of STEREO observations & Universal Time\\
\textsf{CTYPE1}, \textsf{CTYPE2} & Helioprojective (Cartesian) system  & arcsec\\
\textsf{HGLN\_OBS}, \textsf{HGLT\_OBS} & Stonyhurst heliographic longitude \& latitude of the STEREO & degree\\
\textsf{CRLN\_OBS}, \textsf{CRLT\_OBS} & Carrington heliographic longitude \& latitude of the STEREO & degree\\
\textsf{RSUN\_OBS}, \textsf{RSUN} & Observed radius of the Sun in arcsec & arcsec\\
\textsf{RSUN\_REF} & Reference radius of the Sun & meter\\
\textsf{R\_SUN} & Pixel size of the solar radius & pixel\\
\textsf{DSUN\_OBS} & Distance between the center of the Sun and the STEREO & meter\\
\textsf{DSUN\_REF} & Average distance from the Sun to Earth (1 AU) & meter\\
\enddata
\end{deluxetable}

\bibliography{main}{}
\bibliographystyle{aasjournal}

\end{document}